\documentclass[showpacs,twocolumn,aps,preprintnumbers,letterpaper]{revtex4}
\usepackage{amsmath,amssymb}
\usepackage{epsfig}
\usepackage{graphicx}
\usepackage{amsmath}
\usepackage{slashed}
\usepackage{amsfonts}
\usepackage{epstopdf}
\usepackage{color}

\newcommand{\be}{\begin{equation}}
\newcommand{\ee}{\end{equation}}
\newcommand{\bear}{\begin{eqnarray}}
\newcommand{\eear}{\end{eqnarray}}
\newcommand{\ba}{\begin{array}}
\newcommand{\ea}{\end{array}}

\def\be{\begin{eqnarray}}
\def\ee{\end{eqnarray}}
\def\bea{\be}
\def\eea{\ee}

\def\roughly#1{\mathrel{\raise.3ex\hbox{$#1$\kern-.75em%
\lower1ex\hbox{$\sim$}}}}

\begin{document}

\title{Dense Instanton-Dyon Liquid Model: \\Diagrammatics}

\author{Yizhuang Liu, Edward Shuryak   and Ismail Zahed}
\email{yizhuang.liu@stonybrook.edu}
\email{edward.shuryak@stonybrook.edu}
\email{ ismail.zahed@stonybrook.edu}
\affiliation{Department of Physics and Astronomy, Stony Brook University, Stony Brook, New York 11794-3800, USA}


\date{\today}
\begin{abstract}
We revisit the instanton-dyon liquid model in the confined phase by using a non-linear Debye-Huckel (DH)
resummation for the Coulomb interactions induced by the moduli, followed by a cluster expansion. 
The organization is shown to rapidly converge and yields center symmetry
at high density. The dependence of these results on a finite vacuum angle are also discussed.  We also formulate the  
hypernetted chain (HCN) resummation for the dense instanton-dyon liquid and use it to estimate the
liquid pair correlation functions in the DH limit. At very low temperature, the dense limit interpolates between 
chains and rings of instanton-anti-instanton-dyons and a bcc crystal,
with  strong topological and magnetic correlations. 
 \end{abstract}
\pacs{11.25.Tq, 11.15.Kc, 12.38.Lg}


\maketitle

\setcounter{footnote}{0}


\section{Introduction}

This work is a continuation of our earlier studies~\cite{LIU}  of the  gauge 
topology in the confining phase of a theory with the simplest gauge group $SU(2)$.  
We suggested that the confining phase below the transition temperature is an
   ``instanton dyon" (and anti-dyon) plasma which is dense enough
to generate strong screening. The dense plasma is amenable to 
standard mean field methods.

The basic ingredients of the instanton-dyon liquid model are kVBLL  instantons with finite 
holonomies~\cite{KVLL}.  Diakonov and Petrov~\cite{DP,DPX}  have argued that the KvBLL instantons 
split into instanton-dyons in the confined phase below the critical temperature, and recombine above it 
in the deconfined phase. These observations have also been checked numerically~\cite{LARSEN}.
The dissociation of instantons into constituents was advocated originally by Zhitnitsky 
and others~\cite{ARIEL}, and more recently by Unsal and collaborators~\cite{UNSAL} using controlled 
semi-classical approximations. When light quarks are added, center symmetry and chiral symmetry 
are found to be tied~\cite{LIU,SHURYAK1,SHURYAK2,TIN}.

The purpose of this paper is to revisit  the instanton-dyon liquid model without quarks, at low temperature in the center 
symmetric phase, through various many-body re-summations of the Coulomb interactions in the dense limit. In 
particular, we will show that the re-summations provide a specific interpolation between  bion-like correlations
in the dilute phase and mostly screened interactions in the  dense phase.

In section II we briefly review the salient aspects of the instanton-dyon liquid model. We
perform a non-linear Debye-Huckel re-summation of the coulomb interactions stemming from
the moduli space, and combine them with a cluster expansion of the coulomb interactions originating from the
streamlines.  We show that the expansion is rapidly converging and the phase center symmetric already in
the second cluster approximation. In section III we also show how multi-chain and rings can be further re-summed 
beyond the leading clusters and explicit them with some applications. In section IV, we extend our
arguments to a finite vacuum angle $\theta$.  In section V, we discuss  a larger class of resummation
pertinent for dense systems  referred to as  a hypernetted chain re-summation (HCN). In section VI,
we suggest that a melted crystal of instanton-dyons and anti-instanton dyons may provide a
semi-classical description of a Yang-Mills ensemble  at very low temperature.  
Our conclusions are in section VII.  In the Appendix we outline the elements for a future molecular dynamics simulation.

\section{Thermal Yang-Mills}

The chief aspects of the instanton-dyon liquid model have been discussed in~\cite{DP,DPX,LIU} 
to which we refer for more details.
Here, we briefly recall the key elements which will be useful in setting up the statistical Coulomb analysis using
many-body techniques. 
For 2-colors the KvBLL instanton (anti-instanton) splits into  $L,M$ ($\overline L,\overline M$) instanton-dyons
for large holonomies. $M$ carries $(+,+)$ and $L$ carries $(-,-)$ for (electric-magnetic) charges, with
fractional topological charges $\nu$ and $\overline\nu=1-\nu$. The holonomy is fixed by the large x-asymptotics
$\lim_{x\to \infty}\left<A_4^3\right>=2\pi T\nu\tau^3/2$ at fixed temperature $T$. 
In the confined phase  with  $\nu=\frac 12$ and
moderate gauge coupling coupling $\alpha_s\leq 1$ the instanton-dyon actions
$S_L=2\pi\nu/\alpha_s$ and  $S_M=2\pi\overline\nu/\alpha_s$ are still large, justifying their use in a semi-classical description 
of the thermal Yang-Mills phase. Throughout the instanton- and antiinstanton-dyons will carry a finite core size which we will specify below.

A semi-classical ensemble of instanton-antiinstanton-dyons   can be regarded as a statistical
ensemble of semi-classical charges interacting mostly through their moduli space for like instanton- or aniti-instanton-dyons, and
through streamlines for unlike instanton-anti-instanton-dyons. The grand partition function for such an ensemble is of the form
(zero vacuum angle)

\bea
{\cal Z}[T,f]&&\equiv \sum_{[K]}\prod_{i_L=1}^{K_L} \prod_{i_M=1}^{K_M} \prod_{i_{\bar L}=1}^{K_{\bar L}} \prod_{i_{\bar M}=1}^{K_{\bar M}}\nonumber\\
&&\times \int\,\frac{fd^3x_{Li_L}}{K_L!}\frac{fd^3x_{Mi_M}}{K_M!}
\frac{fd^3y_{{\bar L}i_{\bar L}}}{K_{\bar L}!}\frac{fd^3y_{{\bar M}i_{\bar M}}}{K_{\bar M}!}\nonumber\\
&&\times e^{-V(x-y)+{\rm ln\,det}(G[x]G[y])}
\label{SU2}
\eea
The stream-line interactions $V$ are 
large and of order $1/\alpha_s$. They are attractive between like $D\overline D$ and repulsive between unlike $D\overline D$
\cite{LARS}. 
Their relevant form for our considerations will be detailed below.
In contrast, the moduli induced interactions captured in the $(K_L+K_M)^2$ matrix $G[x]$, and in the $(K_{\bar L}+K_{\bar M})^2$ matrix  $G[y]$  are of order $\alpha_s^0$. While the explicit form of these matrices can be found in~\cite{DP,DPX}, it is sufficient
to note here that these induced interactions are attractive between unlike instanton-dyons, and
repulsive between like instanton-dyons. The bare fugacity $f$ will be regarded as an external parameter in what follows. 
Note that in the absence of $V$, ${\cal Z}\rightarrow {\cal Z}_{D}{\cal Z}_{\overline D}$ where each factor can be  exactly re-written in terms of a 3-dimensional effective theory.

 \subsection{Effective action}
 
The streamline interaction part $V$ can be bosonized using the complex fields $b\pm i\sigma$ through standard tricks.
Here $b,\sigma$ refers to the Abelian magnetic and electric potentials stemming from the instanton-dyon charges.  Also, 
each moduli determinant  in (\ref{SU2}) can be fermionized using ghost fields, and the ensuing Coulomb factors bosonized 
using complex $w, \overline w$ fields also through standard tricks as detailed in~\cite{DP,DPX}. The net result of these 
repeated fermionization-bosonization procedures is an exact 3-dimensional effective action (p-space)

 \bea
-S_B[b,\sigma ,w,\bar w]&&=\int d^3p\bigg[\frac{1}{4}(b-i\sigma)V^{-1}(p)(b+i\sigma)\nonumber \\ 
&&\qquad\qquad+\,4\pi \,\left(\nu  f e^{w}+\overline\nu f e^{-w}\right)\nonumber \\ 
&&\qquad\qquad+\,4\pi\, \left(\nu f e^{\bar w}+\overline\nu f e^{-\bar w}\right)\bigg]\nonumber\\
\label{1}
\eea
subject to the constraint from the moduli (x-space)

\be
-\frac{T}{4\pi}\nabla^2(w)+4\pi f\sinh(w)=\frac{T}{4\pi}\nabla^2(b-i\sigma)\nonumber\\
-\frac{T}{4\pi}\nabla^2(\bar w)+4\pi f\sinh(\bar w)=\frac{T}{4\pi}\nabla^2(b+i\sigma)
\label{2}
\ee
(\ref{1}-\ref{2}) allow to re-write exactly the  partition function (\ref{SU2}) in terms of a 3-dimensional effective theory. 
In~\cite{LIU} we have analyzed this partition function using the Debye-Huckel (one-loop) approximation.
Here we will seek a more systematic organization of the dense phase described by (\ref{1}-\ref{2}),
that is more appropriate for the description of the confined phase at low temperature.

\subsection{Cluster expansion}

Our starting point is the linearization of (\ref{2}) around $w=0$ which amounts to the solution

\be
w(p)=\frac{p^2}{p^2+M^2}(b-i\sigma)(p)
\label{3}
\ee
with the squared screening mass $M^2=\frac{16\pi f}{T}$.  Inserting (\ref{3}) into (\ref{1}), we can carry the cluster expansion for
the $4\pi f$ terms by integrating over the $b,\sigma$ fields as the measure is Gaussian in the partition 
function defined now in terms of the 3-dimensional effective action (\ref{1}). The result at second order is

\bea
\frac{\ln Z}{V_3}&&=8\pi f\nonumber\\&&+(4\pi f)^2(\nu^2+{\overline\nu}^2)\int d^3r \,(e^{-V_1(r)}-1)\nonumber \\
&&+(4\pi f)^2(2\nu\overline\nu)\int d^3r\, (e^{-V_2(r)}-1)
\label{4}
\eea
with 

\bea
V_1(p)=-V_2(p)=-\frac{p^4\,V(p)}{(p^2+M^2)^2}
\label{5}
\eea
While the instanton-antiinstanton-dyon interaction is accessible numerically, for simplicity we will use here only
its Coulomb asymptotic form  $V(p)\approx \frac{4\pi C_D}{\alpha_sp^2}$ with $C_D=2$, so that

\be
V_1(r)=-V_2(r)\approx \frac{MC_D}{2\alpha_s}\left(-\frac{2}{Mr}+1\right)e^{-Mr}
\label{6}
\ee
The large r-interaction between the pairs with magnetic charge 0 ($\bar M M$ and $\bar L L$) turns repulsive at large r, 
while that between the pairs with magnetic charge 2  ($\bar M L$ and $\bar L M$) turns  attractive. Remarkably, the sign
of the induced interaction between the pairs in (\ref{6}) is flipped in comparison to the unscreened or bare interaction between
the pairs, a sign of {\it over-screening}.

The chief effect of the moduli constraint (\ref{2}-\ref{3}) is to induce a non-linear Debye-Huckel screening effects between the charged instanton- and anti-instanton-dyons through the Mayer functions $e^{-V_{1,2}}-1$.  This is a re-arrangement of the many-body dynamics that does not assume diluteness. In contrast, the cluster expansion 
in (\ref{4}) is limited to the second cumulant and subsume diluteness in the ensemble of $D,\bar D$ but with non-linear Debye-Huckel
effective interactions. This shortcoming will be addressed later.

\begin{figure}[t]
  \begin{center}
    \includegraphics[width=10cm]{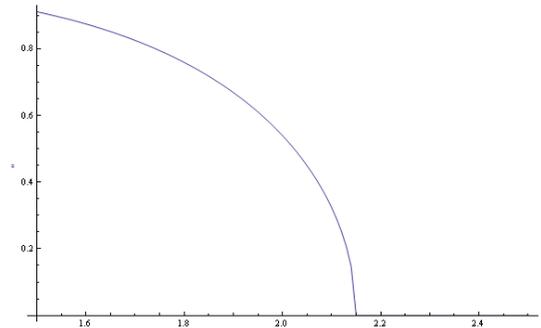}
    \caption{Polyakov line P=$|\cos (\pi \nu)|$ versus m.}
      \label{fig_mpolyakov}
  \end{center}
\end{figure}

For small $r$, we need to set a core for the attractive pair with magnetic charge 2. We choose the core to be
$a=\frac{1}{T}$.  As a result (\ref{4}) plus the perturbative contribution reads


\be
z_b(m,\nu)=&&\frac{\ln Z}{V_3 T^3}-\frac{4\pi^2}{3}\nu^2{\overline\nu}^2\nonumber\\
=&&\frac{m^2}{2}+F(m,\nu)-\frac{4\pi^2}{3}\nu^2{\overline\nu}^2
\label{7}
\ee
with $m=\frac{M}{T}$ and $V(x)=(-2/x+1)e^{-x}$ and 

\be
F(m,\nu)&&=\frac{\pi m}{4}(\nu^2+{\overline\nu}^2)\int_{1}x^2(e^{-\frac{m C_D}{2\alpha_s} V(x)}-1)\nonumber \\
&&+\frac{\pi m}{2}\nu\overline\nu\int_{c1}x^2(e^{\frac{mC_D}{2\alpha_s} V(x)}-1)
\label{8}
\ee
For $C_D\approx 2$ and $ \alpha_s=1$, the transition from a center symmetric (confining) to a center asymmetric (deconfining)
phase occurs for  $m_c\approx$ 2.1, 2.3 for the two choices of the cutoff parameter $c1=1,0$. The choice $c1=0$ corresponds to 
the formal argument presented in~\cite{UNSAL}. In terms of the density  of charged particles $n=8\pi f$,  the transition occurs for $n\approx 2T^3$. For large density, the screening length scales like ${\sqrt{T}}/{\sqrt{n}}$,  while the average separation scales like $1/{n^{\frac{1}{3}}}$.  Our expansion is therefore justified. In Fig.~\ref{fig_mpolyakov} we show the behavior of the Polyakov line versus $m$ for the  cutoff choice $c1=1$.

\section{Open and closed chains}

To go beyond the second cumulant approximation in (\ref{2}) with bare fugacities, 
we will discuss in this section a systematic way for re-summing 
all tree diagrams between the charged particles, and also all ring diagrams with an arbitrary number of trees at the charged
verticies.  One of the chief effect of the 
resummation of all the trees is a re-definition of  the fugacities of the charged
particles as we will show below.

\subsection {Diagrammatics}

\begin{figure}[t]
  \begin{center}
    \includegraphics[width=8cm]{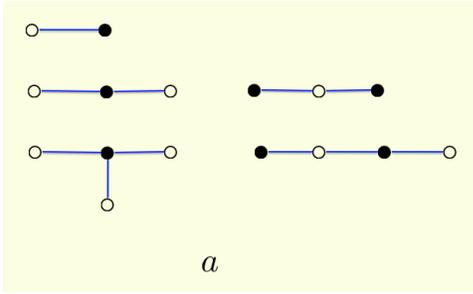}
        \includegraphics[width=8cm]{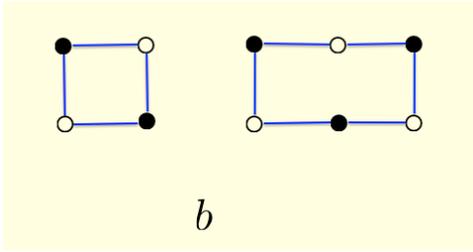}
    \caption{Typical open chain contributions to $\Omega_{\rm tree}$ (a),  and closed chain or ring 
    contributions to $\Omega_{ring}$ (b). $\bullet$ refers to $D=L,M$ and $\circ$ refers to 
    $\bar D=\bar L, \bar M$.}
    \label{fig_tree+ring}
  \end{center}
\end{figure}

A systematic book-keeping procedure for the re-summation of all the trees and the rings
with re-defined fugacities follows from a semi-classical treatment of the Coulomb-like field theory

\be
{\cal  L}=-\frac{1}{2}{\bf \phi}^{T}{ {\bf \cal V}}^{-1}{\bf \phi}+f_1(e^{i\phi_1}+e^{i\phi_3})+f_2(e^{i\phi_2}+e^{i\phi_4})
\label{9}
\ee
with $f_1=4\pi f \nu,f_2=4\pi f \bar \nu $, and the effective fields in 3-dimensions  $\phi=(\phi_1,\phi_2,\phi_3,\phi_4)^T$, 

\be
{\bf \cal V}=\left(\begin{array}{cc}
0&{\bf V}\\
{\bf V}&0
\end{array}\right)\qquad 
{\bf V}=\left(\begin{array}{cc}
F_1& F_2\\
F_2&F_1
\end{array}\right)
\label{10}
\ee
and the Mayer functions $-F_{1,2}=e^{-\beta V_{1,2}}-1$. The $4\times 4$ block-structure follows from the fact that
the statistcal ensemble consists of  4-species of  charged particles 
 $D=L,M$ ($\bullet$) and $\bar D= \bar L,\bar M$ $(\circ)$. The block
off-diagonal character of ${\cal V}$ follows from the fact that the Mayer functions $-F_{1,2}$ resum the non-linear Debye-screening
induced by the moduli between like-instanton-dyons, and are left acting only between unlike $D\bar D$ instanton-antiinstanton-dyons.
It can be checked that (\ref{9}) reproduces all Coulomb diagrams with the correct symmetry and weight factors as Feynman graphs when the  vertices are linked by {\it single lines only}  as illustrated in Figs.~\ref{fig_tree+ring},\ref{fig_56}. 

A re-summation of all trees and rings with arbitrary trees at the vertices amounts to a one-loop expansion 
around the saddle point approximation to (\ref{9}) which is given by

\be
{\bf \phi}_c^T=i{\bf \cal V}(p=0)(f_1e^{i\phi_{1c}},f_2e^{i\phi_{2c}},f_1e^{i\phi_{3c}},e^{i\phi_{4c}})
\label{11}
\ee
Because of symmetry, the solution satisfies $\phi_1=\phi_3$, $\phi_2=\phi_4$. If we define
$\alpha_1=i\phi_{1c},\alpha_2=i\phi_{2c}$ and use the symmetry, then (\ref{11}) reads

\be
\alpha_1=c_1f_1e^{\alpha_1}+c_2f_2e^{\alpha_2}\nonumber\\
\alpha_2=c_1f_2e^{\alpha_2}+c_2f_1e^{\alpha_1}
\label{12}
\ee 
Here 

\be
c_{1,2}=\int d^3x (e^{-\beta V_{1,2}}-1)
\ee
 are the integrated Mayer functions. The saddle point contribution which resums all connected trees yield the pressure

\be
\Omega_{\rm tree}=\frac{\ln Z_{\rm tree}}{V_3}=f_1e^{\alpha_1}(2-\alpha_1)+f_2e^{\alpha_2}(2-\alpha_2)
\label{13}
\ee
with $\alpha_{1,2}$ solutions to the non-linear classical equations (\ref{12}). The resummed rings with arbitrary trees,
follow by expanding (\ref{11}) around the classical solution (\ref{12}) to one-loop. The result is

\bea
\Omega_{\rm ring}=-\frac{1}{2}\int \frac{d^3p}{(2\pi)^3}(\ln (1+\mathbb A)-\mathbb A)
\label{14}
\eea
with in p-space

\bea
\mathbb A=-(\tilde f_1^2+\tilde f_2^2)F_1^2+2\tilde f_1\tilde f_2F_2^2+\tilde f_1^2\tilde f_2^2(F_1^2-F_2^2)^2
\label{15}
\eea
In the special case with $F\equiv F_1\approx -F_2$, the one-loop result  simplifies

\be
\mathbb A=-(\tilde f_1+\tilde f_2)^2F^2
\label{16}
\ee
where  $\tilde f_{1,2}=f_{1,2}e^{\alpha_{1,2}}$ are the tree-modified fugacites.

\subsection{Approximations}

\begin{figure}[t]
  \begin{center}
    \includegraphics[width=8cm]{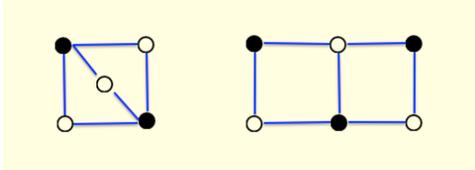}
    \caption{Examples of 2-loop contributions not included in the fugacity redefined 1-loop or ring re-summation.}
        \label{fig_56}
  \end{center}
\end{figure}

The preceding expansion around the small  fugacities follows by seeking the classical solution to
(\ref{12}) in powers $f_{1,2}$ or $\alpha_1\approx c_1f_1+c_2f_2$, $\alpha_2\approx c_1f_2+c_2f_1$.
The tree contributions to the pressure in (\ref{13})   to quadratic order are

\be
\Omega_{\rm tree}\approx 2(f_1+f_2)+c_1(f_1^2+f_2^2)+2c_2f_1f_2
\label{17}
\ee
in agreement with (\ref{4}). For large fugacities $f_{1,2}$  and for $c_1=-c_2=-c<0$,  the solution to (\ref{14}) satisfies
$\alpha_1=-\alpha_2=\alpha$ with $\nu e^{\alpha}=\bar \nu e^{-\alpha}=\sqrt{\bar \nu \nu}$.
As a result, the leading contribution in (\ref{4}) is now changed to

\be
8\pi f \rightarrow 8\pi \tilde f\equiv 8\pi f  \sqrt{4\bar \nu \nu}
\label{19X}
\ee
The resummation of all the trees for large bare fugacities
amount to  {\it dressing} the bare fugacities through $f\rightarrow \tilde f$ in a cluster expansion for the rings with no trees attached
as illustrated in Fig.~\ref{fig_tree+ring}b. Some of the  diagrams not included in the dressed fugacity expansion with ring-diagrams
are illustrated in Fig.~\ref{fig_56} which are of the 2-loop types. The first appear in the 5th cumulant, and the second in the 6th cumulant. So this re-organization resums a large class of diagrams, yet exact up to the 5th cumulant.
Remarkably, in the center symmetric phase with $\nu=\bar\nu=\frac 12$, (\ref{19X}) amounts to the fugacity of
{\it non-interacting} instanton- and anti-instanton-dyons, as all Coulomb interactions from the (linearized)
moduli and the streamlines
average out.

In general, the solution to (\ref{12}) for intermediate fugacities is not emanable analytically. 
One way to go beyond the second cumulant approximation (\ref{17})  at low density is to insert 
the leading solutions $\alpha_1\approx c_1f_1+c_2f_2$,  
$\alpha_2\approx c_1f_2+c_2f_1$ in (\ref{13}) without expanding the exponent, 

\be
\Omega_{\rm tree}\approx 4\pi f\left(\frac 12 +b\right)(2K b+2)e^{-2Kb}+(b\rightarrow -b)\nonumber\\
\label{18}
\ee
where we have set $\nu=\frac{1}{2}+b$, $K=4\pi f c$,  and  noted that  $c_1=-c_2=-c<0$. (\ref{18}) resums all tree 
contributions with charge vertices that include an arbitrary number of 2-body links. (\ref{17}) follows
by expanding the exponents to first order in $f$. We note  that (\ref{18}) has always a maximum 
 at $b=0$ or $\nu=\frac{1}{2}$ for positive $c$ which is center symmetric (confining). This conclusion
 remains unchanged when the ring contributions are added. Indeed, we note that the ring contribution
(\ref{14})  is an increasing function of the  combination   $\tilde f_1+\tilde f_2$  or more specifically

\be
\tilde f_1+\tilde f_2 \approx 8\pi f\left(\frac 12+b \right)e^{-2Kb}+ 8\pi f\left(\frac 12 -b\right)e^{-2Kb}\nonumber\\
\label{19}
\ee
with 

\be
2K=8\pi fc=\frac{2\pi}{m}\int dxx^2(e^{\frac{m C_D}{2\alpha_s}V(x)}-1) 
\label{20}
\ee
using the previous notations.
For $2K>4$ or $cf>\frac{1}{2\pi}$, this combination has a maximum away from 0 and competes 
against the classical contribution towards the center-symmetric solution. For $m<10$ we have $2K< 4$.
The ring contribution preserves center symmetry.

The center symmetric phase can be probed more acuratly by setting $\nu=\frac 12-b$. The semiclassical
equation (\ref{19}) reads

\be
\alpha=-K\left(\frac{1}{2}+b\right)e^{\alpha}+K \left(\frac{1}{2}-b\right)e^{-\alpha}
\label{21}
\ee
At $b=0$ we have $\alpha=0$. We now can solve (\ref{21}) by expanding exactly around $b=0$. Since $\alpha$ is an odd 
function of $b$, we seek a solution to (\ref{21}) using $\alpha=x_1b+x_2b^3+...$, with $x_1$ satisfying

\bea
x_1=&&-2K -K x_1
\label{22}
\eea
Since the leading contribution to the pressure is given by

\be
\frac{\ln Z}{V_3}\approx  4\pi f \nu e^{\alpha}\left(1-\frac{\alpha}{2}\right)
+4\pi f \bar \nu e^{-\alpha}\left(1+\frac{\alpha}{2}\right) +c.c\nonumber\\
\label{23}
\ee
its expanded form to order ${\cal O}(b^4)$ reads

\be
\frac{\ln Z}{V_3}\approx && 8\pi f -8\pi f \frac{2K}{K+1}b^2+{\cal O}(b^4)\nonumber\\
\rightarrow && 8\pi f \left(\frac{1+K\sqrt{4\nu\bar\nu}}{1+K}\right)
\label{24X}
\ee
where the last relation follows after restoring the full $\nu$ dependence. (\ref{24X}) shows that only the
open chains with no tree-like-star insertions contribute to the leading $b^2$ and therefore $\sqrt{\nu\bar \nu}$
in the pressure. Note that (\ref{24X})  is independent of the integrated Mayer function $c$ in $K=4\pi fc$
in the center symmetric phase and/or large fugacities, in agreement with (\ref{19X}).

\section{Finite vacuum angle $\theta$}

At finite vacuum angle $\theta$, the bare fugacities for $\phi_{1,2}$ are now complex and given by
$f_1=4\pi f \nu e^{\frac{i\theta}{2}}$ and $f_2=4\pi f \bar \nu e^{\frac{i\theta}{2}}$, while the bare fugacities 
for $\phi_{3,4}$ are their conjugate $f_{1,2}^{\dagger}$. For $c_1=c_2=-c<0$, we first note 
that the solution to the analogue of the classical equations (\ref{12}) at finite $\theta$ satisfies
$\alpha_{3,4}=\alpha_{1,2}^{\dagger}$,  and  $\alpha_1=-\alpha_2=\alpha$,
with $\alpha$ complex and  satisfying

\be
\alpha=-Ke^{-i\frac{\theta}{2}}\overline\nu e^{\alpha^\dagger}
+Ke^{-i\frac{\theta}{2}}\nu e^{-\alpha^\dagger}
\label{20X}
\ee
The solution for small or large fugacities can be obtained analytically.  We now discuss them sequentially.

\subsection{Large $K$}

For large fugacities or large $K$,
the solution to (\ref{20X}) in leading order gives $e^\alpha=\sqrt{\nu/\overline \nu}$ independently of $K$. In this limit, the summation of all the tree diagrams amount to a dressed fugacity with a leading (dimensionless) pressure 

\be
\frac{\ln Z}{V_3T^3}\rightarrow  \frac {m^2}2\,\sqrt{4\nu\bar \nu}\, \cos \left({\theta}/{2}\right)-\frac{4\pi^2}{3}\nu^2{\overline\nu}^2
\label{X20X}
\ee
with $m^2=\frac {2n}{T^3}$ and
including the perturbative contribution. (\ref{X20X}) resums all the tree cumulant contributions at finite $\theta$ and is to be
compared to (\ref{7}-\ref{8}) with only the second cumulant retained. (\ref{X20X}) implies a transition from the center symmetric
(confined) phase to the center asymmetric (deconfined) phase at a critical temperature

\be
\frac{T_c(\theta)}{T_c(0)}=\left(\cos \left(\frac{\theta+2k\pi}{N_c}\right)\right)^{\frac 13}
\label{T0T}
\ee
with $T^3_c(0)=\frac{12n}{\pi^2}$ for $N_c=2$. 
Although our derivation was for $N_c=2$, our arguments for the re-summation of the trees
extend to any $N_c$. Also, (\ref{X20X}-\ref{T0T}) were derived for $|\theta|<\pi$ in a $2\pi$-branch with $k=0$.  
The general result is multi-branch and $2\pi$-periodic following the substitution $\theta\rightarrow \theta+2k\pi$.
Numerical lattice simulations have established that the transition temperature
$T_c(\theta)$ decreases with $\theta$ as ($k=0$ branch)

\be
\frac{T_c(\theta)}{T_c(0)}=1-R_\theta\theta^2+{\cal O}(\theta^2)
\label{LAT}
\ee
with $R_\theta=0.0175(7)$ for $N_c=3$~\cite{NEGRO},  in good agreement with $R_\theta=1/{6N_c^2}= 0.0185$ from (\ref{T0T}).
Our result (\ref{T0T}) is predictive of the $N_c$ dependence of $R_\theta$ and of the 
higher $\theta$ coefficients, with a cusp at $T_c(\pi)/T_c(0)=1/{2^3}$ at the CP symmetric point. This point is actually
a tri-critical point where the CP breaking first order transition line at $\theta=\pi$ meets the first order transtion cusp from (\ref{T0T}).
Although  (\ref{T0T}) suggests that the CP transition line reduces to a point for $N_c=2$, this conclusion requires further
amendments as it occurs at 0 temperature where the liquid is very dense requiring additional re-summations, some of which will be detailed below.

\subsection{Intermediate $K$}

The onset of the center symmetric phase depends on the details of the
arrangement of the parameters $K,\theta$, as (\ref{X20X}) 
was only established for large $K$ or high density. 
The center symmetric phase can be probed more accuratly for different densities  or $K$ by again setting $\nu=\frac 12-b$
in (\ref{20X}), and solving exactly around $b=0$. The result for the pressure to order ${\cal O}(b^4)$ is

\be
\frac{\ln Z}{V_3}= 8\pi f \cos \frac{\theta}{2}-8\pi f\,2K\frac{K\cos \frac{\theta}{2}-1}{K^2-1}b^2+{\cal O}(b^4)\nonumber\\
\label{24}
\ee
which is seen to reduce to (\ref{24X}) at $\theta=0$. 
At finite vacuum angle $\theta$, the expanded result (\ref{24})   develops a singularity at $K=4\pi fc=1$,
the origin of which requires a more careful analysis. 

In general,  we have $\alpha_1=-\alpha_2$ 
and $\alpha_3=-\alpha_4$. At finite $\theta$, all $\alpha_{1,2,3,4}$
are complex and satisfy the coupled equations

\be
\alpha_1=&&-K e^{-i\frac{\theta}{2}}\nu e^{\alpha_3}
+K e^{-i\frac{\theta}{2}}\overline\nu e^{-\alpha_3}\nonumber\\
\alpha_3=&&-K e^{+i\frac{\theta}{2}}\nu e^{\alpha_1}
+K e^{+i\frac{\theta}{2}}\overline\nu e^{-\alpha_1}
\label{20XX}
\ee
 At small $\theta$, these equations can be analyzed numerically 
by analytically continuing  $\theta\rightarrow -i\theta$, so that

\be
\alpha_1=&&-Ke^{-\frac{\theta}{2}}\left(\frac{1}{2}+b\right)e^{\alpha_3}
+Ke^{-\frac{\theta}{2}}\left(\frac{1}{2}-b\right) e^{-\alpha_3}\nonumber\\
\alpha_3=&&-Ke^{+\frac{\theta}{2}}\left(\frac{1}{2}+b\right)e^{\alpha_1}
+Ke^{+\frac{\theta}{2}}\left(\frac{1}{2}-b\right) e^{-\alpha_1}\nonumber\\
\label{21XX}
\ee
with $\alpha_{1,2,3,4}$ now all real. If we define

\be
f(b,K,\theta,x) =-Ke^{-\frac{\theta}{2}}\left(\frac{1}{2}+b\right)e^{x}
+Ke^{-\frac{\theta}{2}}\left(\frac{1}{2}-b\right) e^{-x}\nonumber\\
\label{21XX1}
\ee
Then  $\alpha_3=x$ satisfies the transcendental equation

\be
f(b,K,-\theta,f(b,K,\theta,x))-x=0
\label{21XX2}
\ee

\begin{figure}[t]
  \begin{center}
    \includegraphics[width=6cm]{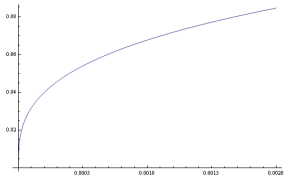}
   \includegraphics[width=6cm]{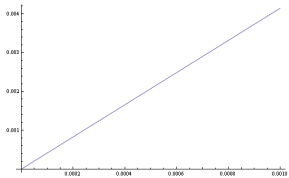}
    \caption{$x(b)$ as a function of $b$ for $K=1$, $\theta=0.1$ (upper) and $K=1.01$, $\theta=0.1$ (lower).}
        \label{fig_SOL}
  \end{center}
\end{figure}

\begin{figure}[t]
  \begin{center}
    \includegraphics[width=10cm]{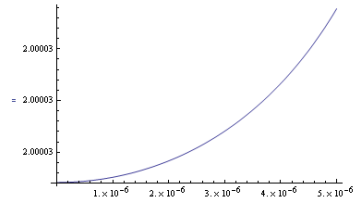}
    \caption{Pressure $\frac{\ln Z}{V_3}$ versus $b$ for $\theta=0.01$ and $K=0.99999$.}
        \label{fig_PX}
  \end{center}
\end{figure}

A numerical analysis of (\ref{21XX2}) reveals  a solution with a 3-branch structure in the parameter space. In the region  $b\ll 1$
around the center symmetric state, it turns out that for  $K$ sufficiently close to 1 but less than 1 
there exists a critical $b_c(K,\theta)$. For $b<b_c(K,\theta)$, the expansion  leading to 
(\ref{24}) is valid. However for $b>b_c(K,\theta)$, the branch which leads to (\ref{24}) no longer
exists, and the solution to (\ref{21XX2}) jumps to a third branch! For $K\ge 1$ and small $b$ only the third branch exists and will lead to the expansion (\ref{24}) for $K>1$. For $K=1$, the solution is more tricky. In Fig.~\ref{fig_SOL} we show the solution
$x(b)$   at $\theta=0.1$ and $K=1$.
In terms of the pressure, it is interesting to see if  a "window" appears for $K=1$. For imaginary $\theta$, we can see a "window" for 
$(1/\cosh(\theta/2))<K<1$ numerically.
Indeed, for $\theta=0.01$ and $K=0.99999$ we show in Fig.~\ref{fig_PX} the
pressure  $\frac{\ln Z}{V_3}$ versus $b$, with no maximum at $b=0$. In contrast,   for $K$ outside the window,
we always have $b=0$ as the maximum, which corresponds to the center symmetric phase. The window disappears  for $\theta=0$. 
Its occurence at finite $\theta$ signals the incompleteness of the tree re-summation for $K$ in the range
$(1/\cos(\theta/2))<K<1$ after analytical continuation.

\section{Hypernetted chains (HNC)}

The static properties of a strongly coupled fluid are usually expressed in terms few-body reduced distribution functions
of which the two-body distribution $g(\vec r_1, \vec r_2)$ or radial distribution $g(r_{12})$ is the standard example. The
radial distribution function describes how the fluid density varies as a function of distance from a reference particle, providing
a link between the microscopic content of the fluid and its macroscopic structure. $g(r_{12})$ can be obtained either from 
simulations using molecular dynamics (see below)  or by solving the Ornstein-Zernicke (OZ) equation~\cite{OZ} subject 
to an additional closure relation. In this section we discuss such a closure in the form of the well-known hypernetted chain re-summation adapted to our dense dyon liquid. For that, we will provide a diagrammatic derivation based on our effective 
field theory (\ref{9}).

\subsection{Diagrammatic derivation}

In the dense instanton-dyon liquid, the radial distribution following from the many-body analysis of (\ref{9}) is a $4\times 4$ matrix
with instanton-dyon entries $g^{ij}(r_{12})$. It is related to the irreducible density 2-point correlation function through

\be
h^{ij}(r)=g^{ij}(r)-1\equiv e^{-\beta {\cal V}^{ij}(r)+\chi^{ij} (r)}-1
\label{HC1}
\ee
where the use of the barometric form in (\ref{HC1}) defines $\chi^{ij}(r)$,
and the $4\times 4$ matrix ${\cal V}$ is given in (\ref{10}).
$\chi^{ij}$ obeys a set of formal matrix equations 

\be
\chi^{ij}=&&\chi^{ij}_a+\chi^{ij}_b\nonumber \\
\chi_a^{ij}=&&c^{il}\,\rho^l\,c^{lj}+c^{il}\,\rho^l\,c^{lm}\,\rho^m\,c^{mj}+....\nonumber\\
c^{ij}=&&h^{ij}-\chi^{ij}_a
\label{HC2}
\ee
where $\rho^{ij}=\rho^i\delta^{ij}$ is a diagonal matrix with species density $\rho^i$. 
We now provide a diagrammatic derivation of  (\ref{HC2}) using the effective formulation (\ref{9}).

\begin{figure}[t]
  \begin{center}
    \includegraphics[width=8cm]{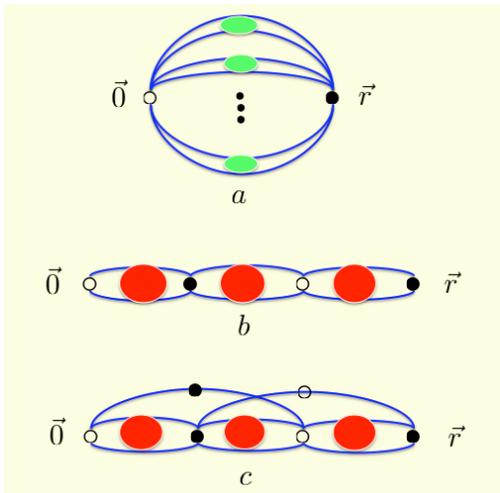}
    \caption{(a) Typical diagrammatic contribution to the pair correlation function $h(r)$ where each hanging ring 
    is $-\beta {\cal V}+\chi$; (b) Typical contribution to $\chi_a$; (c) Typical contribution to $\chi_b$.}
        \label{fig_2pt}
  \end{center}
\end{figure}

The total pair correlation function $h^{ij}$ follows from summing all irreducible graphs with two 
external vertices  fixed between $\vec {0}$ and $\vec r$.  Between these two vertices  we can hang an arbitrary number of 
independent 2-point functions  as illustrated
in~Fig.~\ref{fig_2pt}a. The minimal insertion that cannot be decomposed into such a hanging structure is  denoted
by  $-\beta {\cal V}+\chi$ with $\beta=\frac 1T$. The diagrams contributing to $\chi$ can be separated in type-a and type-b.  
Type-a have at least one cutting point, i.e. a vertex that one can cut to split the diagram into two disconnected pieces
as illustrated in~Fig.~\ref{fig_2pt}b, while type-b have none as illustrated
in~Fig.~\ref{fig_2pt}c. For type-a, we can further  count by enumerating the number  of cutting points and define a summation over all possible 2-point diagrams that can be put between two nearest cutting points as  $c(r)$,
which defines the direct correlation function. It is readily seen that $c=h-\chi_a$.  
With these definitions  in mind, simple diagrammatic arguments yield (\ref{HC2}).
The hypernetted chain approximation (HNC) amounts to setting $\chi_b=0$. In this case, (\ref{HC2}) can be cast in the more standard form

\be
h^{ij}=&&c^{ij}+ c^{ik}\rho^k\star h^{kj}\nonumber\\ 
c^{ij}=&& -\beta {\cal V}^{ij} +h^{ij}- \ln (1+h^{ij})
\label{HNC}
\ee  
where  $\star$ means convolution in x-space. The first of these equations is known as the Ornstein-Zernicke (OZ) equation,
while the  second equation as the HNC closure condition. The interaction energy per 3-volume and therefore the pressure can be
re-constructed using  the pair correlation function, for instance

\be
\frac{\cal E}{V_3}= \left(\frac{2N}{V_3}\right)\,\frac 12\,\sum_{i,j}\int d^3r\,\beta {\cal V}^{ij}(r)h^{ij}(r)
\ee

\subsection{Linear and non-linear DH approximations}

The linear Debye-Huckel (DH) approximation follows by performing one iteration in the OZ equation 
with the initial condition $h=0$ or $c\approx-\beta {\cal V}$,  to obtain formally in p-space

\be
h_{DH}=\frac{-\beta {\cal V}}{1+\beta \rho {\cal V}}
\label{DH1}
\ee
For the instanton-dyon ensemble we have $\rho=\rho_1=\rho_2= M^2T/8$
 and $V_1=-V_2=-V(p)=-\frac{8\pi p^2}{(p^2+M^2)^2}$ in ${\cal V}$,  so that

\be
h_{DH}&&=\frac{\beta V }{1-(2\beta \rho V)^2}
\left(\begin{array}{cc}
{2\beta \rho V}& {1}\\ {1}& {2\beta \rho V}
\end{array}\right)\otimes (1-\sigma_1)\nonumber\\
\label{DH1X}
\ee
Here $\sigma_1$  is a Pauli matrix. 
(\ref{DH1X}) defines two independent pair correlation functions in p-space

\be
&&h_{MM}=h_{LL}=-h_{ML}=\frac{2\rho \,(8\pi \beta)^2 p^4}{(p^2+M^2)^4-(16\pi\beta \rho)^2p^4}\nonumber\\
&&h_{M \bar M}=h_{L\bar L}=-h_{M\bar L}=\frac{(8\pi \beta)\, p^2(p^2+M^2)^2}{(p^2+M^2)^4-(16\pi \beta \rho)^2p^4}\nonumber\\
\label{DHX2}
\ee
For $\rho= {M^2T}/8$ the denominator 

\be
(p^2+M^2)^4-4\pi^2p^4M^4
\label{POLE}
\ee
is negative for $p>\frac M{\sqrt{2\pi-1}}\approx \frac M2$. The spatial cutoff $a=\frac 1T$ used earlier, translates to 
a p-cutoff of $T$. Since $M\approx 2T$, the negative  range is physically not relevant. These observations are similar 
to the ones encountered in the DH analysis of  the electric and magnetic correlation functions in~\cite{LIU} (first reference).

The HNC equations (\ref{HNC}) allow to go beyond the DH approximation in the dense ensemble, but requires a numerical
calculation. Here, we only mention that a simple non-linear correction to the DH result follows from (\ref{HNC}) by retaining
the leading correction to the direct correlation function, namely 
$c^{ij}\approx -\beta {\cal V}^{ij}+\frac 12 (h^{ij})^2$, and use it to iterate the OZ equation after the substitution $h\rightarrow h_{DH}$.
The net effect is a non-linear correction to the DH result (\ref{DH1}) in p-space

\be
h_{DH2}=\frac{-\beta {\cal V}+\frac 12 h_{DH}^2}{1+\rho (\beta {\cal V}-\frac 12  h_{DH}^2)}
\label{DH2}
\ee

\section{Instanton-dyon crystal}

At even higher fugacity or density, the instanton- and anti-instanton dyons are expected to crystalize.
A typical bcc cubic crystal arrangement with low energy is  illustrated in Fig.~\ref{fig_crystal}.
Recall that the re-summed $M\bar M$ interactions and $L\bar L$ interactions are repulsive, while the $L M$ and $\bar L M$
interactions are attractive. In the bcc crystal structure, we note that the nearest neighbor $LM$ vertices are close to 
an instanton configuration, while their alternate nearest neighbors $\bar L M$ vertices are close to a magnetically charge 2
bion. We will refer to this as  crystal duality. We note that  holographic dyonic-crystals composed only of $L,M$ in salt-like
or popcorn-like crystal configurations were suggested in~\cite{RHO} for a holographic description of dense matter.

The instanton- and anti-instanton-dyons considered throughout are the lightest of a Kaluza-Klein tower
with higher winding numbers which carry larger actions (more massive). 
We expect them to crystallize following a similar pattern, albeit with 
higher windings. We expect this tower of 3-dimensional crystal arrangements along the extra winding direction to be dual to 
a 4-dimensional crystal arrangement of monopoles and anti-monopoles (or instantons and anti-instantons by crystal
duality), using  the  Poisson duality suggested in~\cite{UNSAL}.  Remarkably, the resulting 4-dimensional 
and semiclassical description  at very low temperature, can be either described 
as instanton-like (topologically charged) or monopole-like (magnetically charged) as the two descriptions are tied
by crystal duality.

The crystal is
an idealized description of the strongly coupled and dense phase as both the low temperature and the quantum fluctuations cause 
it to melt. The melted form of Fig.~\ref{fig_crystal} resembles an ionic liquid  with 4 species of ions with strong local order.
This  semi-classical description of the Yang-Mills state at very low temperature
appears to reconcile the instanton liquid model without confinement,
with the t$^\prime$Hooft-Mandelstam proposal with confinement. In the former,
the low temperature thermal state is composed of a liquid of instanton and anti-instantons, while in the latter 
it is a  superfluid  of monopoles and anti-monopoles with bions as precursors~\cite{UNSAL}. The dual descriptions  allow for a  center symmetric thermal state with both strong and local topological and magnetic correlations.

\begin{figure}[t]
  \begin{center}
    \includegraphics[width=8cm]{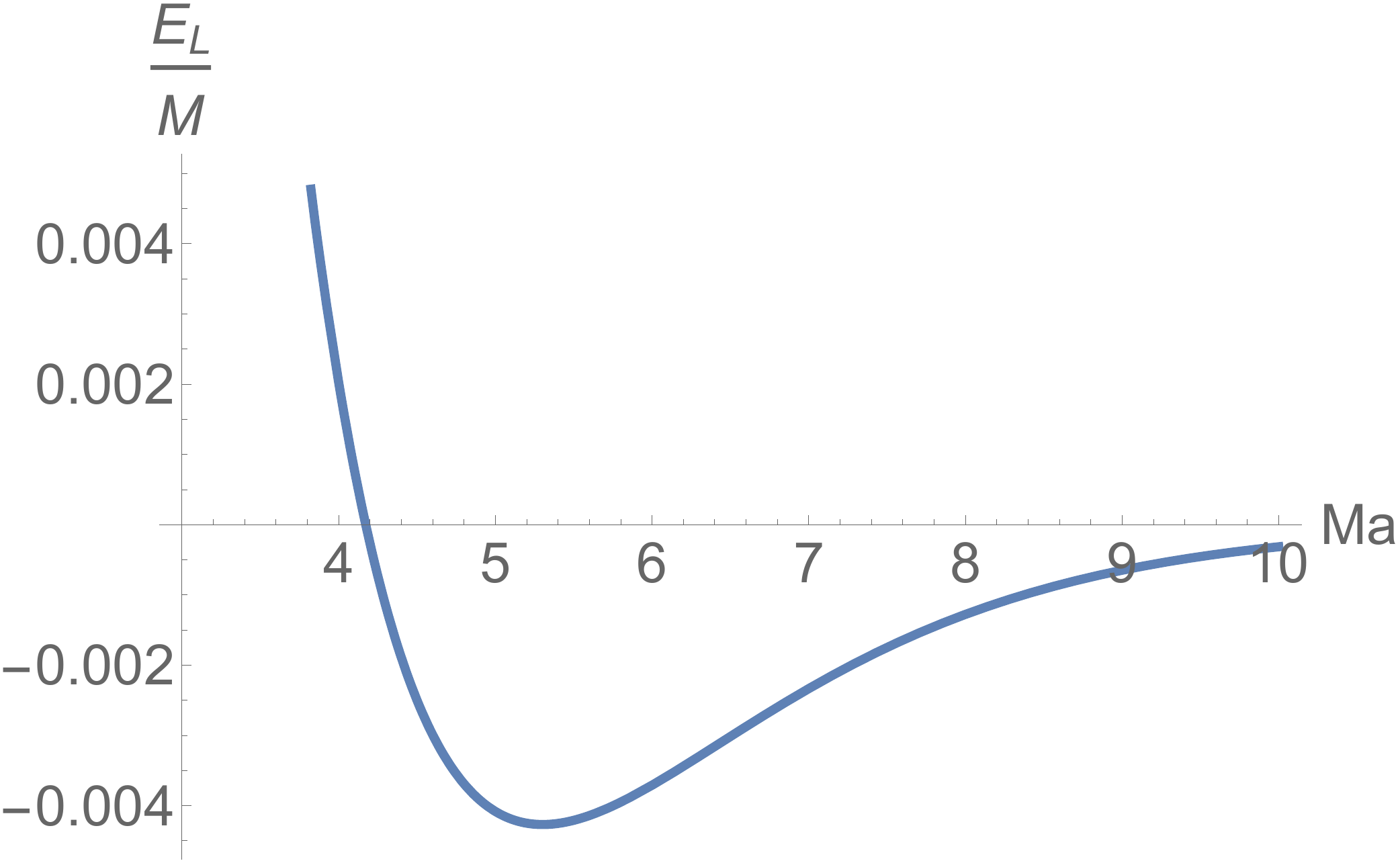}
    \caption{Crystal energy for the bcc arrangement $E_L/M$ versus $Ma$ as given in (\ref{bccx}) with $\alpha_s=1$.}
            \label{fig_ELX}
  \end{center}
\end{figure}

\subsection{Crystal energy}

\begin{figure}[t]
  \begin{center}
    \includegraphics[width=8cm]{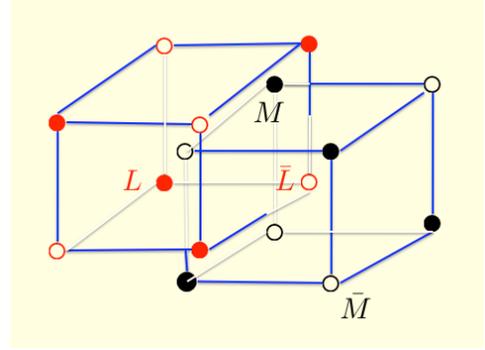}
    \caption{3-dimensional bcc crystal composed of the instanton- and antinstanton-dyons with the lowest winding, for 2 colors.}
        \label{fig_crystal}
  \end{center}
\end{figure}

To assess the crystal contribution to the pressure at high density, we first evaluate the  interaction energy for the crystal structure in Fig.~\ref{fig_crystal}. Consider the $L$ instanton-dyon sitting in the center of the M cell. The interaction summation within the
L-lattice reads

\be
2E_L=\sum_{n_1,n_2,n_3\ne 0 } (1-(-1)^{n_1+n_2+n_3})V(\vec{r}_{n_1,n_2,n_3})
\label{C1X}
\ee
The mutual interaction between the L- and M-lattice  is

\be
-2E_{ML}= &&\sum_{n_1,n_2,n_3} (1-(-1)^{n_1+n_2+n_3})\nonumber\\
&&\times V\left(\frac 12 {\vec{r}_{111}}+\vec{r}_{n_1,n_2,n_3}\right)
\label{C2X}
\ee
In momentum space, these sums can be cast using the dual lattice $\vec b_{n}=\frac{2\pi}{a}\vec{n}$, using the identity

\be
\sum_{a_{n}}e^{ip\cdot a_n}=\sum_{n}\delta(p-b_n)
\ee
The results are

\be
2E_{L}=&&\sum_{n}\left(V(b_n)-V\left(b_n+\frac{\pi}{a}(1,1,1)\right)\right)\nonumber\\
-2E_{ML}=&&\sum_{n}(-1)^{n}\left(V(b_n)-V\left(b_n+\frac{\pi}{a}(1,1,1)\right)\right)\nonumber\\
\label{CP}
\ee
where we made use of  

\be
e^{ib_n\cdot\frac{r_{111}}{2}}=(-1)^{n_1+n_2+n_3}=(-1)^n
\ee
Both  the x-space sums (\ref{C1X}-\ref{C2X}) and the p-space sums (\ref{CP}) can only be carried numerically.
However, we note that the x-sum is converging exponentially and can be approximated by the leading contribution 
involving only the nearest neighbors,

\be
E_L\equiv ME_L(\tilde a) \approx  M\left(6V(\tilde a)-4V\left(\frac{\sqrt{3}\tilde a}{2}\right)\right)
\label{bccx}
\ee
with $V(x)=\frac 1{\alpha_s}(-\frac{2}{x}+1)e^{-x}$ from (\ref{6}) with $C_D=2$. Here we have set $\tilde a=Ma$, 
with $M^2=\frac{2n}T$ and  $n=8\pi f$. Note that the total energy of the crystal is extensive

\be
E(N,M)\approx 2NME_{L}\left(\tilde a\equiv Ma=\left(\frac{M^3V_3}{2N}\right)^{\frac{1}{3}}\right) 
\ee
In Fig.~\ref{fig_ELX} we show the behavior of (\ref{bccx})  for $\alpha_s=1$. The bcc configuration
is bound for $\tilde a =Ma\approx 5$, but the binding energy is very small $E_L/M\approx -0.004$. 

\subsection{Disordered crystal pressure}

\begin{figure}[t]
  \begin{center}
    \includegraphics[width=8cm]{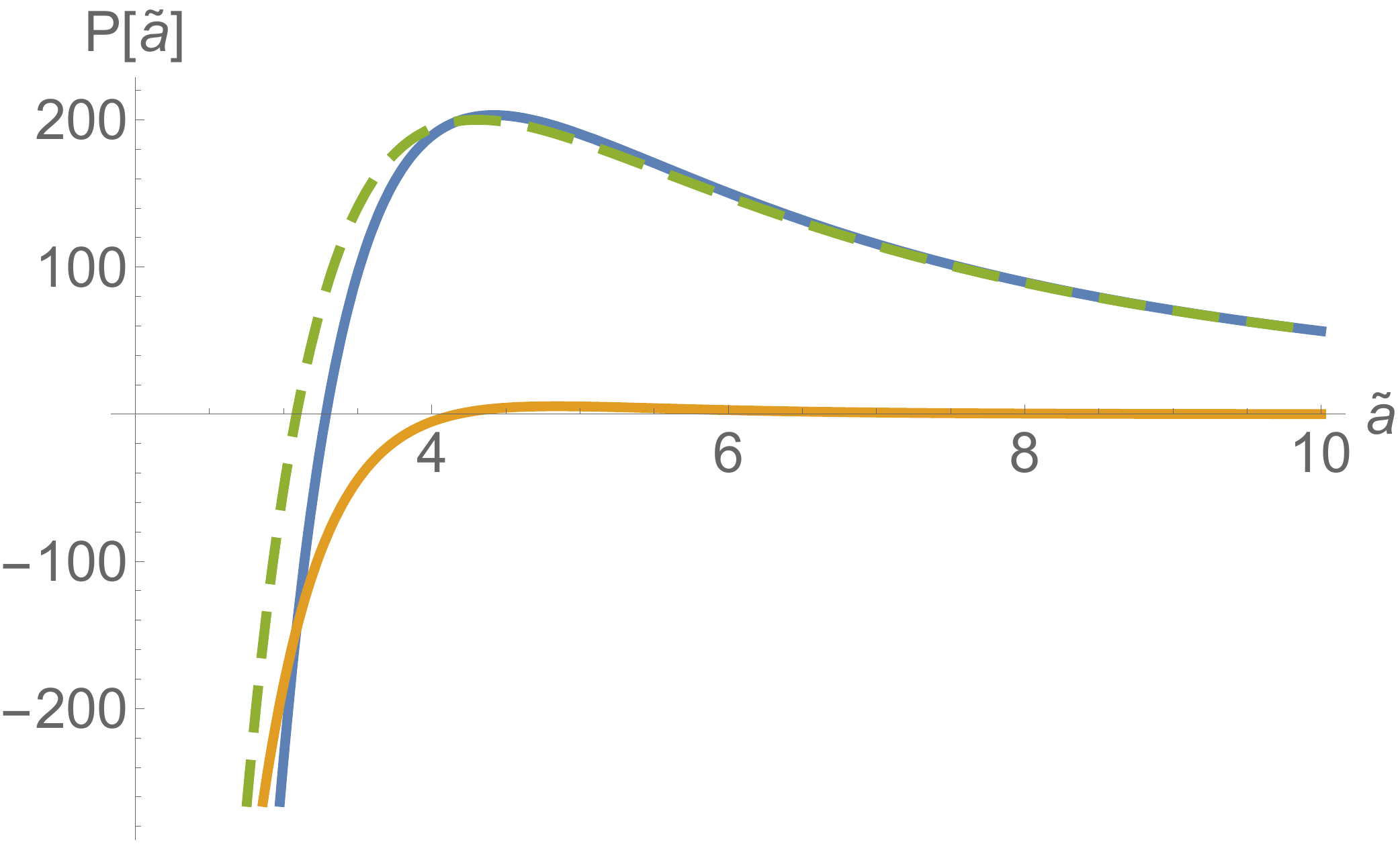}
    \caption{Pressure (\ref{bcc1}) versus $\tilde a$ for $m=20$ and $\nu=\frac 12$ upper-solid-curve. The separate
    contributions from the crystal  (first term in (\ref{bcc1})) is shown as the lower-solid-curve,  and the entropy of the re-summed trees
    (second term in (\ref{bcc1}))  is shown as the dashed-curve.}
        \label{fig_px}
  \end{center}
\end{figure}

The pressure for a disordered crystal follows from the corresponding partition function

\be
\Omega_{\rm bcc}(\tilde a) =\sum_{N}\frac{(V_3 2\pi \tilde f)^{4N}}{(N!)^4}e^{-\frac {2NM}T E_L(\tilde a)}
\ee
where we used the quantum and dressed fugacity $2\pi\tilde f=2\pi f\sqrt{4\nu \bar \nu}$ from  (\ref{19X}). In the large N-limit,
the pressure ${\cal P}={{\rm ln}\Omega_{\rm bcc}}/{V_3}$ can be cast in the form

\be
\frac{{\cal P}(\tilde a)}{T^3}\approx 
-\frac{m^4}{{\tilde a}^3}E_L({\tilde a})+\frac {2m^3}{{\tilde a}^3}\,\left(1+{\rm ln}\left(\frac{{\tilde a}^3\sqrt{\nu\bar\nu}}{2m}\right)\right)
\label{bcc1}
\ee
with $m=M/T$ (the ratio of the screening mass to the temperature). The first contribution in (\ref{bcc1}) is the crystal energy,
and the second contribution is the entropy of the competing trees at large density as discussed in IVA. 
For large $m$ (very low temperature)
the pressure is dominated by the crystal contribution, while for small $m$ (intermediate  temperature) the
pressure is dominated by the entropy of the trees. 
In Fig.~\ref{fig_px} we show the behavior of the pressure ${\cal P}(\tilde a)$
versus $\tilde a$ for $m=20$ for the center symmetric case with $\nu=\frac 12$ upper-solid-curve,
while the crystal contribution is shown as the lower-solid-curve, and the tree contribution as the
dashed-curve. The pressure is maximum at

\be
\frac{{\cal P}_{\rm max}}{T^3}=&&\frac {m^2}2\sqrt{4\nu\bar\nu}e^{-\frac m2(E_L(\tilde a_\star)-{\tilde a_\star}^3E_L^\prime(\tilde a_\star))}
\nonumber\\&&\times \left(1-\frac{m {\tilde a_\star}^3}2 E_L^\prime(\tilde a_\star\right)
\label{bcc2}
\ee
with $\tilde a_\star$ solution to the transcendental equation

\be
\frac{{\tilde a}^3_\star  \sqrt{\nu\bar\nu}}{2m }=   e^{\frac m2(E_L(\tilde a_\star)-{\tilde a_\star}^3E_L^\prime(\tilde a_\star))}
\label{bcc3}
\ee
If we were to assume $E_L$ fixed at the crystal minimum and constant
as in Fig.~\ref{fig_ELX},  i.e $E_{L\rm min}\approx -0.004$, then (\ref{bcc2}) simplifies

\be
\frac{{\cal P}_{\rm max}}{T^3}\rightarrow  \frac {m^2}2 \sqrt{4\nu\bar\nu}\,e^{-\frac m2 {E_{L\rm min}}}
\label{bcc}
\ee
which is seen to interpolate between the re-summed tree contribution (\ref{X20X}) at small $m$ (intermediate temperature) 
and the crystal at large $m$ (very low temperature). Due to the small binding energy of the crystal shown in Fig~\ref{fig_ELX}, 
the crystal contribution takes over only when $\frac m2$ is large   or very high density (very low temperature). This is confirmed numerically. Note that in both (\ref{bcc1}) and (\ref{bcc}) the ratio $\frac m2$ plays the role of the Coulomb factor. 
It is rather large with $\frac m2=500$  for the onset of the crystal.

\section{Conclusions}

We have provided a many-body analysis of the instanton-dyon liquid model in the center symmetric
phase. The starting point of the analysis was a linearization of the moduli interactions beween like
instanton-dyons $DD$ and anti-instanton-dyons ($\bar D\bar D$),  followed by a cluster expansion. 
This re-organization of the many-body physics was shown to be captured exactly by a 3-dimensional 
effective theory between charged particles. A semi-classical treatment of this effective theory amounts
to re-summing the tree contributions in the form of effective fugacities, while the 1-loop correction 
amounts to re-summing all ring or chain diagrams with effective fugacities. The tree or chain contributions are found to 
yield a center symmetric phase even at finite vacuum angle. They are dominant in the range $1\leq\frac m2\leq 10$.

At very low temperature or large fugacities, an even larger class of diagrams need to be re-summed. 
In this vein we have carried the HNC re-summation as is commonly used for dense and charged liquids,
and used it to estimate  the pair correlation function around the DH approximation
in the dense instanton-dyon liquid. The very low temperature
phase is argued to be a melted bcc crystal with strong local topological and magnetic correlations. 
 A simple description of the thermodynamics of an ensemble composed of trees and bcc crystals show that the tree-like 
 contributions are dominant for most temperatures, with the exception of the very low temperature regime where the crystal arrangement is more favorable owing to its very small binding.  
 To better understand the range of validity of the present diagrammatic results,  it will be important to carry a full molecular dynamics 
 calculation for comparison. This point will be addressed next.

\section{\label{acknowledgements} Acknowledgements}

 This work was supported in part by the U.S. Department of Energy under Contract No.
DE-FG-88ER40388.


\section{Appendix: Molecular dynamics}

 (\ref{SU2}) describes a 4-species ensemble of charged  particles  in 3-spatial dimensions.
 For fixed fugacity, the statistical ensemble described by (\ref{SU2}) can be recovered from ensembles
 of classically evolved electrically and magnetically charged particles in 3-dimensions by sampling over 
 random initial conditions. All the particles carry 
equal (dimensionless) mass  $m_D=f^{\frac 23}/2\pi T^2$, and move classically following the Newtonian paths fixed  by
 
 \bea
 m_D\,\ddot{x}_n=&&-\frac{\partial}{\partial x_n}\sum_{[i, \bar i]} \left(V(x_i-y_{\bar i})-{\rm ln}{\rm det}G[x_i]\right)\nonumber\\
m_D\, \ddot{y}_{\bar n}=&&-\frac{\partial}{\partial y_{\bar n}}\sum_{[i, \bar i]} \left(V(x_i-y_{\bar i})-{\rm ln}{\rm det}G[y_{\bar i}]\right)
 \label{26}
 \eea
 The first contribution is the Coulomb force stemming from the streamline potential, while the second contribution
 is the Coulomb force following  from the moduli. The latter is of the form ${\rm Tr}(G^{-1}\partial_nG)$. It requires inverting $G$ 
at each time step, which may prove   numerically costly for molecular dynamics (MD) simulations.  It also requires that 
${\rm det}\,G\neq 0$ for the inversion to be valid. For this the role of the initial conditions is important~\cite{LIU} (first reference).

To remedy some of these shortcomings, we recall
that a linearization of the effects induced by the moduli interactions amounts to  non-linear Debye-Huckel interactions between the pair $D,\bar D$ as captured by (\ref{9}), leading simpler  MD equations

\be
\left(\begin{array}{cc}
{\bf m}_D\,{\ddot{ x}}_D\\
{\bf m}_{\bar D}\,{\ddot{x}}_{\bar D}
\end{array}\right)=
\left(\begin{array}{cc}
-\frac{\partial}{\partial{x}_D}\\
-\frac{\partial}{\partial{x}_{\bar D}}
\end{array}\right)\,
\sum_{i=1,2; \underline{D}\,\underline{\bar D}} {V}_{i}({x}_{\underline D}-{x}_{\underline {\bar D}}; \theta)
\label{27}
\ee
Here the mass ${\bf m}_D$ is for  the pair $D=L,M$ and ${\bf m}_{\bar D}={\bf m}_D^\dagger$ for the pair $\bar D=\bar L, \bar M$ at finite vacuum angle

 \be
{\bf m}_D\equiv  (m_L,m_M)=m_D\,e^{\frac {i\theta}3}\,(\nu^{\frac 23}, {\bar\nu}^{\frac 23})
 \ee
The potentials in (\ref{27}) generalize (\ref{6}) to finite vacuuum angle

\be
&&V_1(r; \theta)=-V_2(r; \theta )=\nonumber\\&&-\frac{C_D}{2\alpha_s\,r}\left(e^{-Mre^{i\theta/4}+i\theta}+e^{-Mre^{-i\theta/4}-i\theta}\right)
\label{28}
\ee
Note that the MD analysis of (\ref{27}) for $\theta\neq 0$ is more challenging as it generates complex trajectories.

 \vfill

\end{document}